\documentclass[journal]{IEEEtran}
\usepackage{cite}

\ifCLASSINFOpdf
\else
\fi
\usepackage{amsmath}
\usepackage{dblfloatfix}
\usepackage{url}

\usepackage{booktabs}
\usepackage{pgfplots}
\usetikzlibrary{positioning}
\usetikzlibrary{calc}
\usetikzlibrary{shapes.multipart}
\usepackage{mathtools}
\usepackage[caption=false]{subfig}

\begin{document}
%
\title{Perceptual Loss with Recognition Model for Single-Channel Enhancement and Robust ASR}
%
%
%

\author{Peter~Plantinga,~\IEEEmembership{Student~Member,~IEEE,}
        Deblin~Bagchi,~\IEEEmembership{Member,~IEEE,}
        and~Eric~Fosler-Lussier,~\IEEEmembership{Senior~Member,~IEEE}
\thanks{P. Plantinga, D. Bagchi, and E. Fosler-Lussier were all with the Ohio State University in Columbus Ohio for this work. e-mail: plantinga.1@osu.edu}
}

%
%

\markboth{Journal of \LaTeX\ Class Files,~Vol.~14, No.~8, August~2015}%
{Shell \MakeLowercase{\textit{et al.}}: Bare Demo of IEEEtran.cls for IEEE Journals}
%



\maketitle

\begin{abstract}
Single-channel speech enhancement approaches do not always improve
automatic recognition rates in the presence of noise, because
they can introduce distortions unhelpful for recognition. Following
a trend towards end-to-end training of sequential neural network
models, several research
groups have addressed this problem with joint training of front-end
enhancement module with back-end recognition module. While this
approach ensures enhancement outputs are helpful for recognition,
the enhancement model can overfit to the training data,
weakening the recognition model in the presence of unseen noise.
To address this, we used a pre-trained acoustic model to generate
a perceptual loss that makes speech enchancement more aware of
the phonetic properties of the signal. This approach keeps
some benefits of joint training, while
alleviating the overfitting problem. Experiments on Voicebank + DEMAND
dataset for enhancement show that this approach achieves a new
state of the art for some objective enhancement scores.
In combination with distortion-independent training, 
our approach gets a WER of 2.80\% on the test set, which is
more than 20\% relative better 
recognition performance than joint training, and 14\% relative better than
distortion-independent mask training.
\end{abstract}


%
\IEEEpeerreviewmaketitle

\section{Introduction}
\label{sec:intro}
%
%
%
%
\IEEEPARstart{M}{any} applications for recorded speech need to be
intelligible to both human listeners and artificial listeners. For example,
teleconference systems may need to automatically improve speech quality to
assist human users of the system,
and produce automatic captions at the same time for record-keeping or
for users with hearing difficulties. Real-time translation services
are another example, since better automatic recognition will improve
translation quality, at the same time as reducing noise in recordings.

Speech enhancement systems have been shown to improve intelligibility for
human listeners, especially if they have hearing loss~\cite{zhao2016dnn}.
A reasonable hypothesis would be that speech enhancement systems already
improve automatic speech recognition (ASR) systems without the need for special
loss functions or other similar techniques.
However, research has shown that this is not always the case,
since enhancement systems can sometimes introduce distortions that are
not helpful for recognition~\cite{narayanan2014investigation,wang2019bridging,tan2020improving}.

A popular approach to improving the cooperation between an enhancement system
and an automatic recognition system is to train them jointly, using the recognition
objective to ensure enhancement outputs are good for recognition~\cite{gao2015joint,wang2016joint,ravanelli2017network,xu2019joint,menne2019investigation}. This
follows a trend in machine learning of training multiple sequential models
with the end-goal objective, i.e. end-to-end training \cite{heigold2016end,heymann2017beamnet,zeyer2018improved}.
From this body of work we can tell that using joint training
can be an effective strategy for recognition system improvement.

One limitation of joint training in the context of noise-robust ASR is that the enhancement module can overfit the training data, reducing the ability of the recognition module to handle distortions. This problem is known as the ``distortion problem'' and was first noted by Wang, Tan, and Wang in~\cite{wang2019bridging}. In this paper, the authors note that enhancement modules are not likely to distort the data they are trained on. If a recognition model back-end is trained on the same dataset as its enhancement model front-end, it never learns to adapt to distortions produced by the enhancement model. The solution proposed by the authors is to train the recognition model independently from the enhancement model and add additional noises to the noisy inputs during recognizer training. This forces the enhancement model to perform inference while the recognition model is being trained,
giving the recognition model a chance to adapt to distortions.

While this approach does mitigate the distortion problem, it also loses some of the benefits of joint training. The enhancement model is not trained with any knowledge of recognition objectives, and myopically focuses on energy differences. This has several limitations. As an example, Turian and Henry have demonstrated 
that spectral measures of distance do not capture pitch differences, which is important for tasks such as tonal language recognition~\cite{turian2020m}.
Another example of the limitations of spectral measures is that
they can miss low-energy
phonemes due to over-emphasis on energy differences~\cite{plantinga2020phonetic}.

\begin{figure}
\tikzstyle{label} = [fill=white,opacity=0.7,text opacity=0.9,font=\small]
\centering
\begin{tikzpicture}
	
	\node[rectangle,draw] (noisy) at (0,0) {\includegraphics[width=3.5cm,height=2cm]{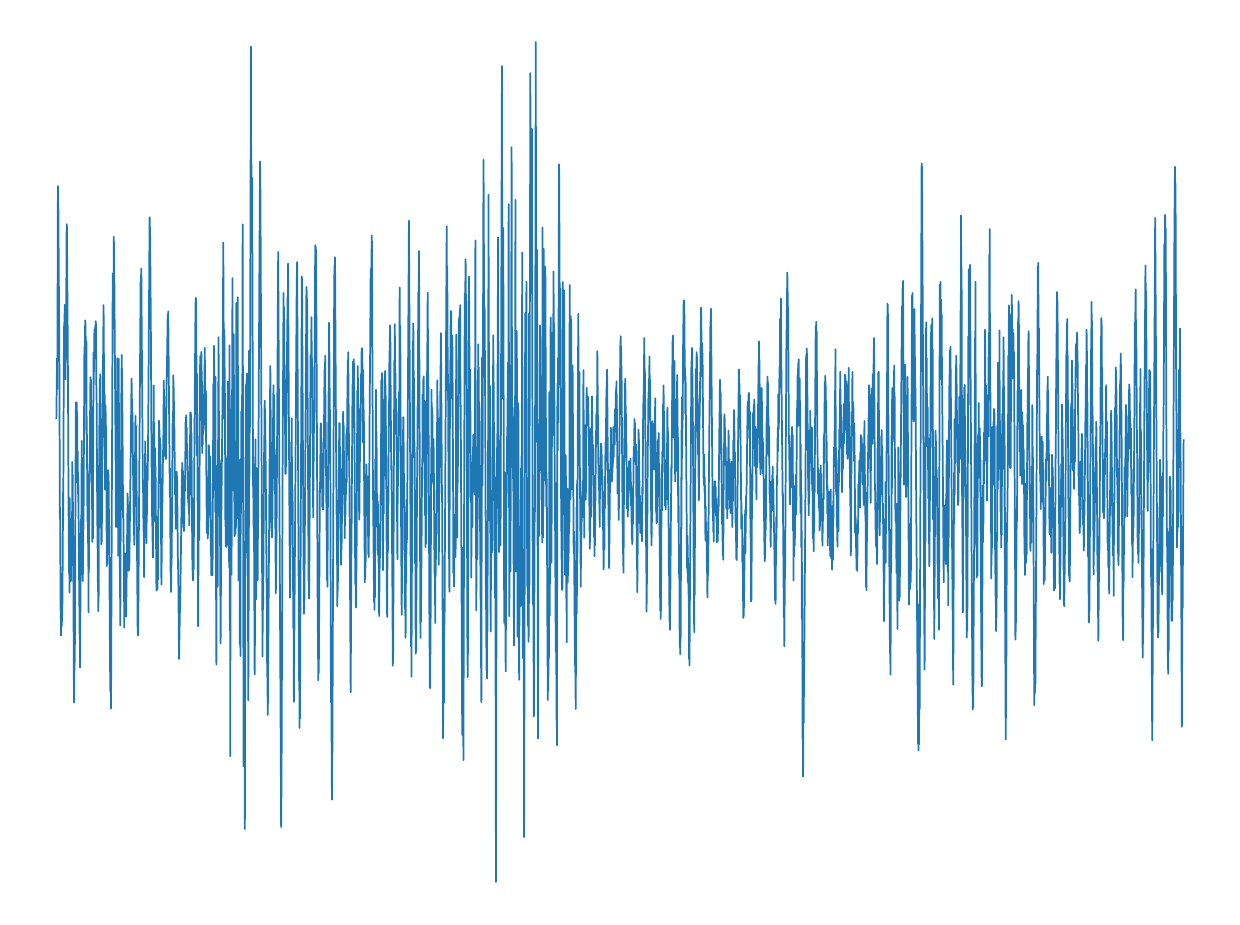}};
    \node[label,below=0.1em of noisy.north] (noisy_label) {noisy signal};
    
    \node[rectangle,draw,below=3.8em of noisy] (denoised) {\includegraphics[width=3.5cm,height=2cm]{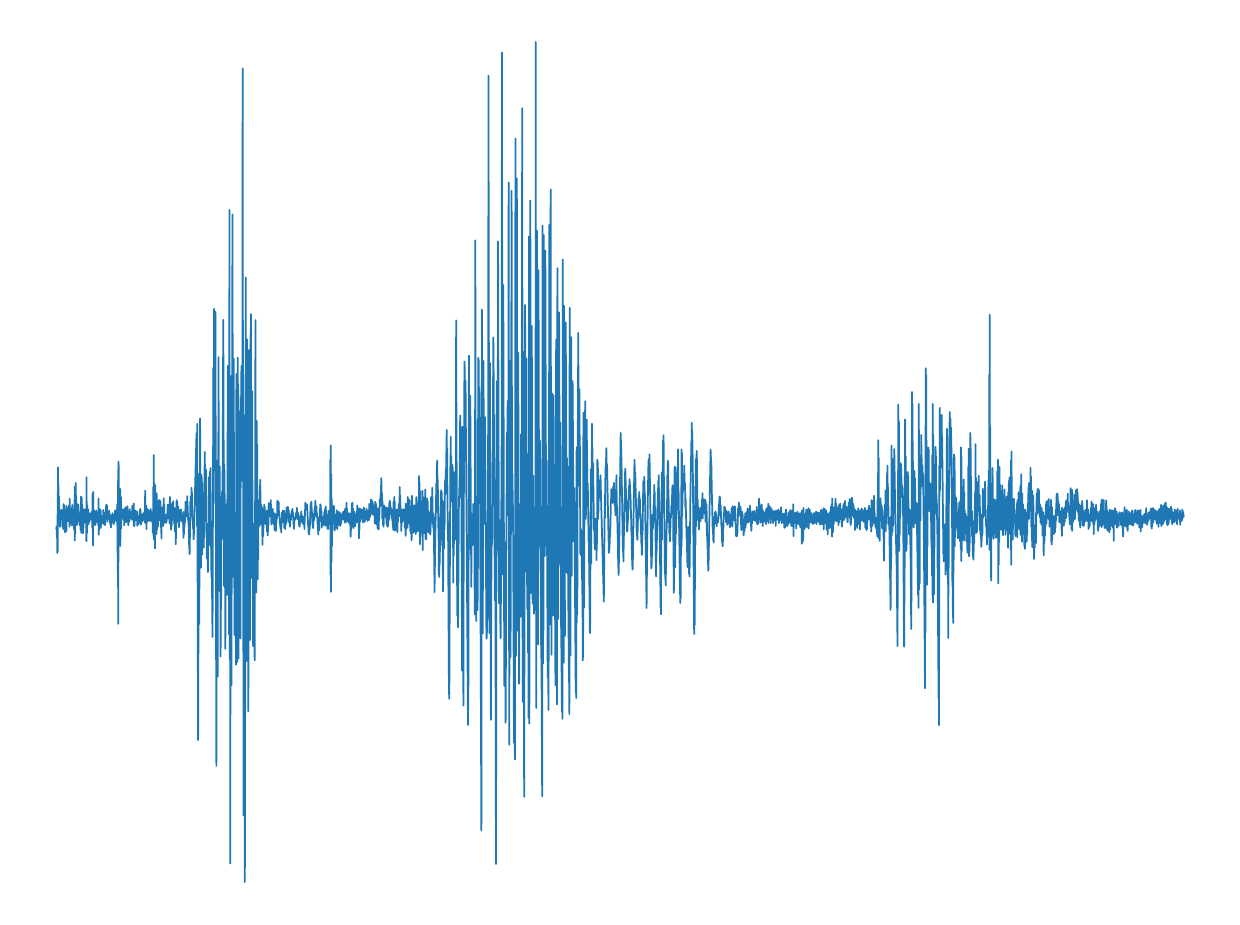}};
    \node[label,below=0.1em of denoised.north] (denoised_label) {denoised signal};
    
    \node[rectangle,draw,below=3.8em of denoised] (senones) {\includegraphics[width=3.5cm,height=2cm,angle=180,origin=c]{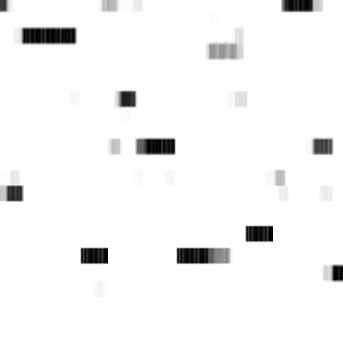}};
    \node[label,below=0.1em of senones.north] (post) {posterior};
    \draw[decoration={brace,raise=5pt,amplitude=0.3cm,aspect=0.5},decorate](noisy.north west) -- node[above=0.5cm,pos=0.5] {\Large{\begin{tabular}{c}Step 1: \\ Enhancement\end{tabular}}} (noisy.north east);
    

    \node[rectangle,draw,right=0.5cm of noisy] (denoised2) {\includegraphics[width=3.5cm,height=2cm]{figures/denoised_waveform.pdf}};
    \node[label,below=0.1em of denoised2.north] (despec2)  {denoised signal};
    \node[rectangle,draw,below=3.8em of denoised2] (senones2) {\includegraphics[width=3.5cm,height=2cm,angle=180,origin=c]{figures/posteriorgram}};
        \node[label,below=0.1em of senones2.north] (post2) {posterior};
    \node[rectangle,draw,below=3.8em of senones2] (words) {\includegraphics[width=3.5cm,height=2cm]{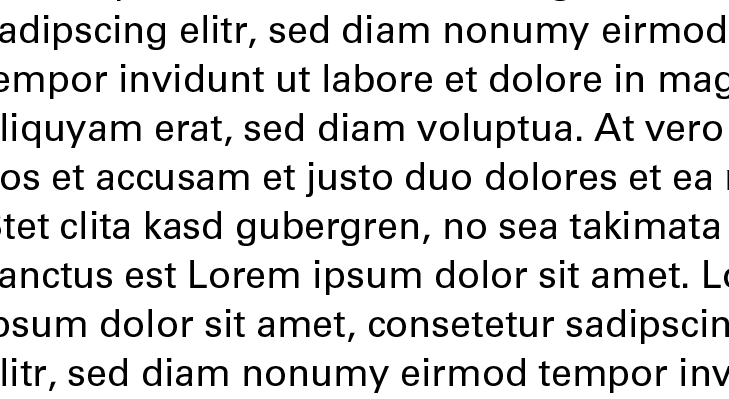}};
    \node[label,below=0.1em of words.north] (word) {recognized words};
    \draw[decoration={brace,raise=5pt,amplitude=0.3cm,aspect=0.5},decorate](denoised2.north west) -- node[above=0.5cm,pos=0.5] {\Large{\begin{tabular}{c}Step 2: \\ Recognition\end{tabular}}} (denoised2.north east);
    
    \draw[-stealth,ultra thick] (noisy) -- node[left] {\textbf{\begin{tabular}{c}Signal \\ denoiser\end{tabular}}} (denoised);
    
    
    \draw[-stealth,ultra thick,color=red,dashed] ([xshift=0.3cm] denoised.north) -- node[right=-0.1em] {\textbf{\begin{tabular}{c}Perceptual \\loss\end{tabular}}} ([xshift=0.3cm] noisy.south);
    
    \draw[-stealth,thick] (denoised) -- node[left] {\begin{tabular}{c}Phoneme \\ recognizer\end{tabular}} (senones);
    
    \draw[-stealth,thick,color=red,dashed] ([xshift=0.3cm] senones.north) -- node[right] {\begin{tabular}{c}Perceptual \\ loss\end{tabular}} ([xshift=0.3cm] denoised.south);
    
    \draw[-stealth,ultra thick] (denoised2) -- node[right] {\textbf{\begin{tabular}{c}Acoustic\\modeling\end{tabular}}} (senones2);
    
    \draw[-stealth,thick] (senones2) -- node[right=0.2cm] {Decoding}  (words);
    
\end{tikzpicture}
\captionsetup{margin=1cm}
\caption{Basic setup for training with perceptual loss. This setup proves effective for improving
both enhancement quality and recognition results. Bold indicates model training.}
\label{fig:pipeline}
\end{figure}

We have recently proposed an approach that mitigates the distortion problem,
while maintaining some of the advantages of joint training, depicted in Fig.~\ref{fig:pipeline}. Our approach uses a
recognition model pre-trained with phoneme targets and clean input 
speech to generate a phonetic perceptual loss\footnote{We originally framed this loss with the name ``mimic loss'' as it teaches the enhancement module to induce acoustic model behavior similar to its behavior under clean speech; however, with the rise of other perceptual loss work in the field, it seems best to reframe this loss in this newer light.} for improved enhancement
training~\cite{bagchi2018spectral,plantinga2018exploration,plantinga2020phonetic}.
This approach preserves modularity,
and achieved state-of-the-art recognition scores on the CHiME-2
challenge~\cite{vincent2013second} for systems using the default language model.

In this work, we report an extended in-depth analysis of our results from previous
experiments~\cite{bagchi2018spectral,plantinga2018exploration} showing that
perceptual loss improves recognition of low-energy phonemes over than high-energy phonemes. In addition,
we report a follow-up experiment that changed the perceptual model to use only
local information. This experiment matched what is now the state-of-the-art
recognition score on CHiME-2.

In addition to our analysis, we show the generality of our approach by developing a perceptual loss recipe  using the SpeechBrain
toolkit~\cite{ravanelli2021speechbrain}. The new experiments serve several purposes. First, we directly compare
for the first time two techniques for noise-robust ASR: perceptual training and joint training.
Second, we report both speech recognition and speech enhancement results a publicly-available dataset in keeping with SpeechBrain philosophy of repeatable results, filling
a void in the robust ASR literature. Third, we make the recipe and pre-trained models
available as part of the toolkit so that anyone can use or train models with our perceptual loss.

The results of our experiments with the SpeechBrain framework show that we
achieve state-of-the-art enhancement performance
on the Voicebank + DEMAND dataset for
enhancement~\cite{valentini2017noisy}. At the same time,
passing the enhanced data to an unadapted noisy recognition
model achieves similar performance as the jointly trained enhancement
and recognition systems. Finally, a recognition model that was adapted to
distortions from the enhancement model reached a 2.80\% WER on
the Voicebank + DEMAND test set, a 20\% relative improvement over
joint training and a 14\% relative improvement over
distortion-independent masking.

\section{Background}
\label{sec:background}

Approaches for joint training of recognition module and enhancement module
can be quite varied. Joint training was first proposed by Gao, Du, Dai, and Lee in~\cite{gao2015joint}
using a spectral mapping model to achieve good results on the Aurora4 task.

This approach can also apply to masking-based enhancement, as proposed by
Wang and Wang in~\cite{wang2016joint}. The authors show that joint
training with masking, with the addition of features that are noise-robust
by design, are able to achieve good
recognition rates on the CHiME-2 challenge dataset~\cite{vincent2013second}.

A more sophisticated joint training and communication scheme was proposed
by Ravanelli, Brakel, Omologo, and Bengio in ~\cite{ravanelli2017network}.
The authors noted that communication between front-end model and back-end
model could be improved by a multi-stage network that explicitly
passes information from stage to stage.

There are in fact several denoising strategies that can be jointly
trained with the roboust ASR system. Joint training has been done with a
masking-based beamforming model~\cite{xu2019joint}. Joint training
has also been done with a Weiner filter approach~\cite{menne2019investigation}.

Related to joint training is the technique of perceptual loss training, which
uses a secondary pre-trained model in order to provide feedback to the primary
model. This technique was first defined by Johnson, Alahi, and Fei-Fei 
in~\cite{johnson2016perceptual}. Perceptual loss was shown to correlate well
with human judgements by Zhang et al. in~\cite{zhang2018unreasonable}.

The use of a perceptual loss via a phoneme classifier was first proposed by
Oord et al. for the task of speech synthesis~\cite{oord2018parallel}. The
authors used a WaveNet-like model trained to recognize phonemes from raw
audio. In addition, they found the best results using the Euclidean
distance between the gram matrices, as in style loss~\cite{johnson2016perceptual}.

In addition to speech synthesis, perceptual loss is rapidly gaining popularity
for speech enhancement. Parallel to our earliest work on this topic, Germain, Chen,
and Koltun introduced perceptual loss for enhancement (which the authors call
``deep feature loss'') which is used on the time-domain
signal~\cite{germain2018speech}. This approach uses
a perceptual model based on acoustic scene classification, whereas our approach
uses an acoustic model as the perceptual model.

Another perceptual loss that was proposed after our work is that of Yao and Al-Dahle,
who propose a ``Dynamic Perceptual Loss'' for speech enhancement~\cite{yao2019coarse}. 
This loss compares the activations of a discriminator during GAN training, in addition
to the binary output loss used in traditional GAN training.

One more perceptual loss used for speech enhancement was named by the authors
``Phone-Fortified Perceptual Loss''~\cite{hsieh2020improving}.
This loss consists of using as a perceptual
model the wav2vec 2.0 model~\cite{baevski2020wav2vec}. Although the motivation
and description of this loss by the authors sounds similar to using a phoneme
classification perceptual model,
in practice the perceptual model they use, called wav2vec, is trained in an
unsupervised manner and has no explicit access to phonetic information.

Another work by Kataria, Villalba, and Dehak
tested a variety of perceptual losses~\cite{kataria2021perceptual}
using large-scale pre-trained models for different tasks. The full list of tasks
includes: acoustic event
classification, acoustic modeling, speaker embedding, speech emotion classification,
and self-supervised feature extraction (the authors use PASE+~\cite{ravanelli2020multi}, and wav2vec 2.0~\cite{baevski2020wav2vec}). The enhancement model used by the authors is
a strong enhancement architecture based on the Conformer
model~\cite{gulati2020conformer}. Surprisingly, they found no significant improvement
from any perceptual loss other than the acoustic event classification model.

Especially relevent to our work is that the authors found no improvement using a
large-scale pretrained acoustic model. The particular acoustic model that they
chose is DeepSpeech 2~\cite{amodei2016end}, which uses character targets. Our 
own experiments suggest that phonemic targets are much more effective. In addition,
the authors use an RNN layer as part of their perceptual model which we found to be
harmful. More details about the settings we found to correlate with good enhancement
performance can be found in Section~\ref{sec:voicebank_procedure}.

By contrast to the work evaluating perceptual losses for enhancement, our work evaluates
the same model both for enhancement as well as robust ASR. Our experiments show that
perceptual loss can recover some of the benefits of joint training without suffering
from the distortion problem.

\section{CHiME-2 Experiments}
\label{sec:chime2_experiments}

\begin{figure}
\centering
\tikzstyle{block} = [rectangle, draw, minimum height=3.5em, minimum width=5em,align=center,fill=blue!10]
\tikzstyle{conv} = [block, fill=white, minimum width=7em, rounded corners]
\begin{tikzpicture}
	\node (input) {Noisy frames};
    \node[block, below=1em of input] (block1) {128 filter \\ block};
    \node[block, below=1em of block1] (block2) {128 filter \\ block};
    \node[block, below=1em of block2] (block3) {256 filter \\ block};
    \node[block, below=1em of block3] (block4) {256 filter \\ block};
    \node[block, below=1em of block4] (full) {Fully \\ connected};
    \node[below=1em of full] (output) {Denoised frame};
    
    \draw[-stealth,thick] (input) -- (block1);
    \draw[-stealth,thick] (block1) -- (block2);
    \draw[-stealth,thick] (block2) -- (block3);
    \draw[-stealth,thick] (block3) -- (block4);
    \draw[-stealth,thick] (block4) -- (full);
    \draw[-stealth,thick] (full) -- (output);
    
    \node[block, right=3em of block3, minimum height=22.8em, minimum width=10em] (outset) {};
    
    \draw[thick] (block3.north east) -- (outset.north west);
    \draw[thick] (block3.south east) -- (outset.south west);
    
    \node[conv, below=1em of outset.north] (conv1) {Size: $3 \times 3$ \\ Stride: $2 \times 2^\ast$};
    \node[conv, below=2em of conv1] (conv2) {Size: $3 \times 3$ \\ Stride: $1 \times 1$};
    \node[conv, below=1em of conv2] (conv3) {Size: $3 \times 3$ \\ Stride: $1 \times 1$};
    \node[conv, below=1em of conv3] (squeeze) {Squeeze \& \\ Excitation$^\dagger$};
    \node[circle, draw, below=1em of squeeze, fill=white] (add) {+};
    
    \draw[-stealth,thick] (outset.north) -- (conv1);
    \draw[-stealth,thick] (conv1) -- (conv2);
    \draw[-stealth,thick] (conv2) -- (conv3);
    \draw[-stealth,thick] (conv3) -- (squeeze);
    \draw[-stealth,thick] (squeeze) -- (add);
    \draw[-stealth,thick] (add) -- (outset.south);
    
    \coordinate[below=0.9em of conv1] (a);
    \coordinate[right=4.3em of a] (b);
    \coordinate[right=4.3em of add.center] (c);
	\draw[-stealth,thick, rounded corners] (a) -- (b) -- (c) -- (add);
\end{tikzpicture}

\caption{Our Wide ResNet-like model. $\ast$The stride is only 1 in the time dimension for our Voicebank experiments. $\dagger$The squeeze \& excitation layer was not
present in our CHiME-2 work but was added for the Voicebank work.}
\label{fig:resnet}
\end{figure}
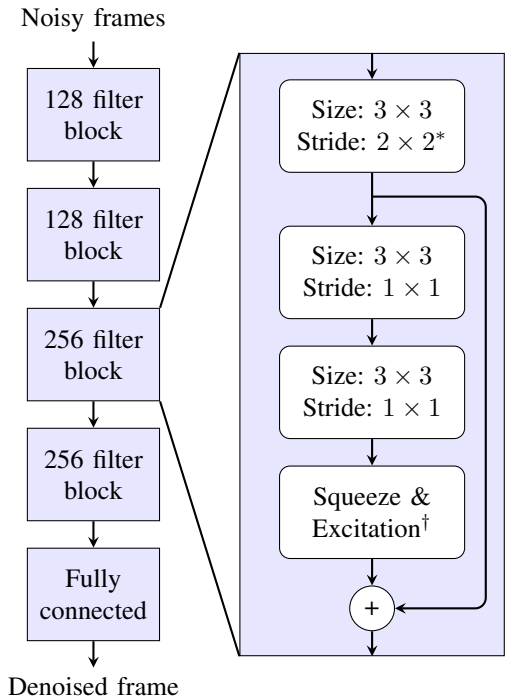

We repeat our experimental settings for our prior CHiME-2 experiments
here in order to provide the background for an in-depth analysis. In addition,
we report one follow up experiment that did not make it into the
original report~\cite{plantinga2018exploration}.

Our experimental procedure followed a three step process:

\begin{enumerate}
    \item Pretrain an acoustic model using clean speech to serve as a perceptual model.
    \item Train an enhancement model using perceptual loss.
    \item Use the off-the-shelf Kaldi CHiME-2 recipe to evaluate recognizability.
\end{enumerate}

\begin{table*}[!htbp]

\centering
\caption{WER for our systems relative to state-of-the-art on CHiME-2 broken down across different SNRs.\\   *Result reported here for the first time.}
\label{tab:chime2}
\begin{tabular}{|ccc|cccccc|c|}
\toprule
Enhancement model & Perceptual model & Recognition model & -6 dB & -3 dB & 0 dB & 3 dB & 6 dB & 9 dB & Avg\\
\midrule
- & - & DNN-HMM & 29.9 & 21.5 & 17.7 & 15.0 & 10.7 & 9.4 & 17.4 \\
DNN & - & DNN-HMM & 29.4 & 19.2 & 16.1 & 11.8 & 10.3 & 9.3 & 16.0 \\
DNN & DNN & DNN-HMM & 26.2 & 18.7 & 14.1 & 11.1 & 8.9 & 7.7 & 14.4 \\
DNN & WRBN & DNN-HMM & 25.6 & 17.8 & 13.8 & 10.5 & 8.8 & 7.8 & 14.0 \\
DNN* & Wide ResNet* & DNN-HMM & 26.3 & 18.5 & 13.8 & 10.4 & 8.6 & 8.0 & 14.2 \\
Wide ResNet & - & DNN-HMM & 18.7 & 13.9 & 10.7 & 8.1 & 6.9 & 6.2 & 10.8 \\
Wide ResNet & DNN & DNN-HMM & 19.3 & 14.3 & 10.2 & 7.4 & 6.2 & 5.4 & 10.5 \\
Wide ResNet & WRBN & DNN-HMM & 17.0 & 11.8 & 8.8 & 7.1 & 6.0 & 5.2 & 9.3 \\
Wide ResNet* & Wide ResNet* & DNN-HMM & 15.2 & 10.9 & \textbf{8.3} & \textbf{6.7} & \textbf{5.8} & \textbf{5.2} & \textbf{8.7} \\
\midrule
GRN (2019)~\cite{wang2019bridging} & - & WRBN & \textbf{14.8} & \textbf{10.0} & 9.0 & 6.8 & 6.3 & 5.5 & \textbf{8.7} \\
\bottomrule
\end{tabular}
\end{table*}

\begin{table}
\caption{Effect of changing the output layer of the perceptual model on recognition performance.}
\label{tab:layers}
\begin{center}
\begin{tabular}{|l|c|c|}
\toprule
Perceptual output & value of $\alpha$ & WER \\
\midrule
Just L1 loss & - & 16.0 \\
First layer & 0.5 & 15.0 \\
Second layer & 1.0 & 15.1 \\
Third layer & 1.0 & 14.7 \\
Fourth layer & 1.0 & 14.9 \\
Fifth layer & 1.0 & 14.6 \\
Sixth layer & 2.0 & 14.4 \\
\bottomrule
\end{tabular}
\end{center}
\end{table}

For our first step, we generated senone alignments using standard Kaldi tools.
These alignments match every frame in the spectral features with a class out
of a set of 1999 senones. The senones are generated by careful clustering
of the set of triphone states to a reasonably-sized set, which means they encode
important phonetic information.

Our perceptual models were trained with cross-entropy loss against aligned senone
targets. Our features were standard 257-dimensional spectral features
with 25ms windows and 10ms hop. These are identical in form to the features generated
by the enhancement model, allowing easy passage of the gradient to the enhancement model.

We experimented with three different acoustic models. Our previously reported experiments
covered two important types of model: a frame + context 6-layer DNN
model~\cite{bagchi2018spectral},
as well as an utterance-wise good performing Wide Residual BiLSTM Network (WRBN) model~\cite{jahn2016wide}.

We ran additional experiments that have not yet been reported using a third acoustic model.
Our third model still takes only frame + context input,
but provides deeper feedback. This third model is a Wide-ResNet-like model, nearly
identical to the one we used for enhancement. The model is shown in Fig.~\ref{fig:resnet}.

In the second step of training, we add perceptual loss to a standard L2 loss on the
spectral magnitudes.
This perceptual loss is generated by passing both clean and denoised spectral features
to the pretrained acoustic model, and computing the L2 distance of the respective outputs. We used a scaling parameter $\alpha$ to keep a similar magnitude for the
two losses.

The enhancement models take 5 frames of past and future context, and they output a single denoised frame. The first enhancement model is a 2-layer 2048-node DNN network with leaky ReLU activations. The
second enhancement model is quite similar to Wide ResNet~\cite{zagoruyko2016wide}.
The full network has 4 blocks of three convolutional layers, where the first layer
is used for downsampling (stride 2), and the other two layers compute a residual that is
added to the result. The first two blocks have 128 filters, and the second two have 256 filters. Two fully-connected layers are inserted before the output layer.
The enhancement models were more fully described in the original
work~\cite{plantinga2018exploration}.

After training the enhancement model, we passed the enhanced features to an off-the-shelf Kaldi~\cite{povey2011kaldi} recipe for recognition on the CHiME-2 dataset. This recipe trains a HMM-DNN based system for recognition using cross-entropy loss against the senone targets. The model is further fine-tuned using sMBR loss. This same recipe gets 2-3\% WER on the WSJ, demonstrating its recognition ability. We did not change any part of the recipe, including the default trigram language model. Our changes were restricted to the input features.

\section{CHiME-2 Results and analysis}
\label{sec:chime2_analysis}

\begin{figure*}
\centering
\subfloat[Noisy utterance]{%
  \includegraphics[width=6cm]{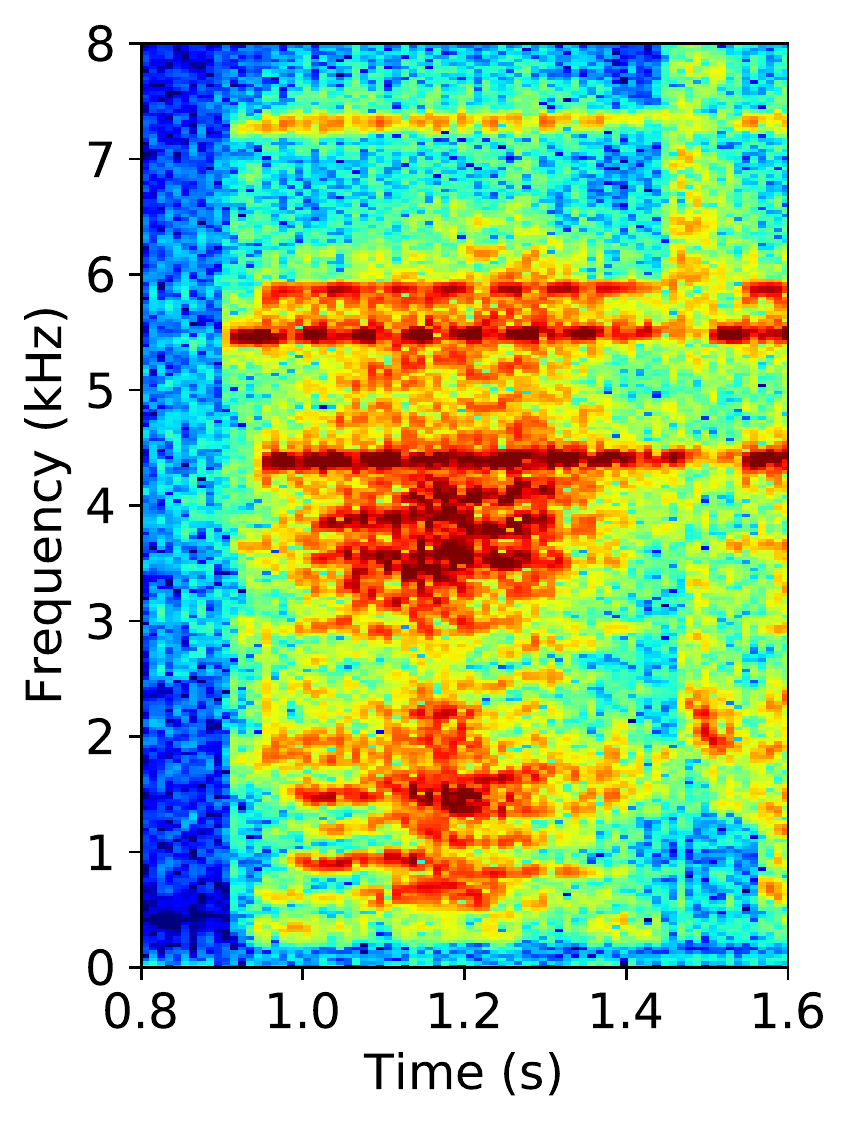}%
  \label{fig:evaluation:noisy}%
}%
\subfloat[DNN mapper w/ L1 loss]{%
  \includegraphics[width=6cm]{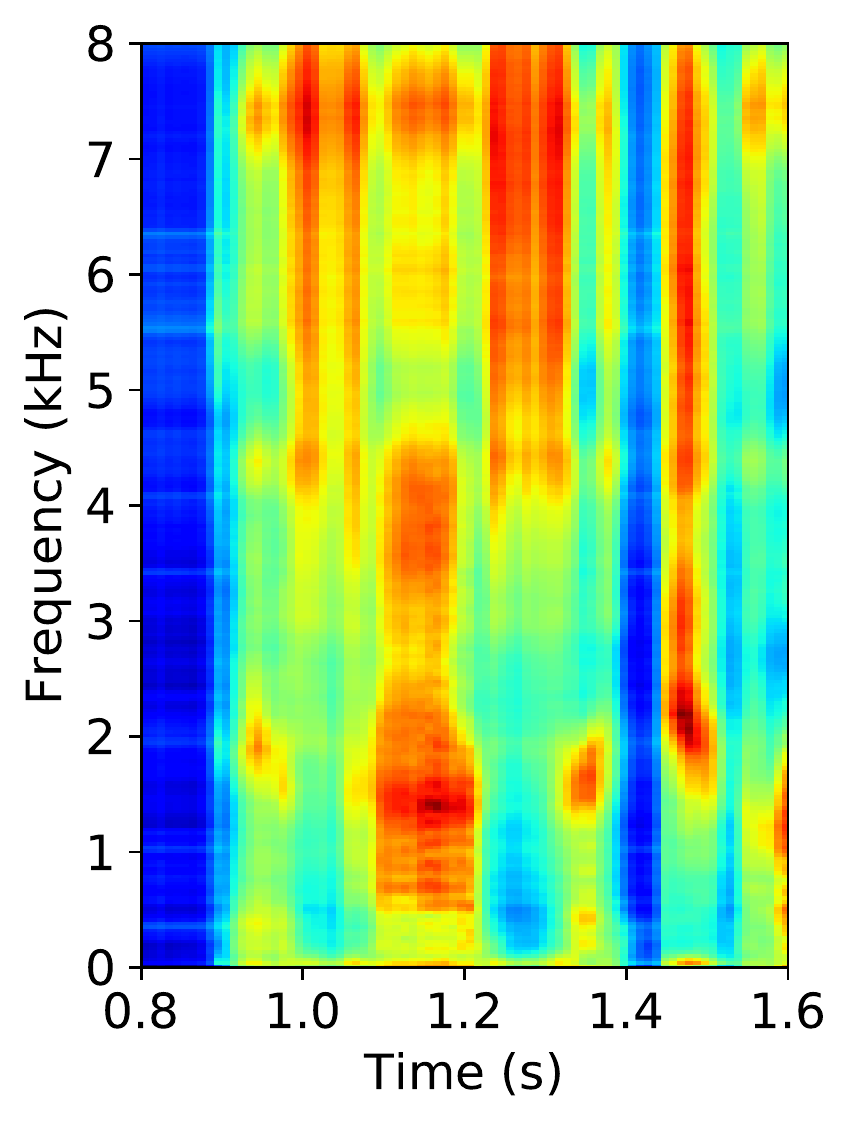}%
  \label{fig:evaluation:dnnfidelity}%
}%
\subfloat[DNN + perceptual DNN]{%
  \includegraphics[width=6cm]{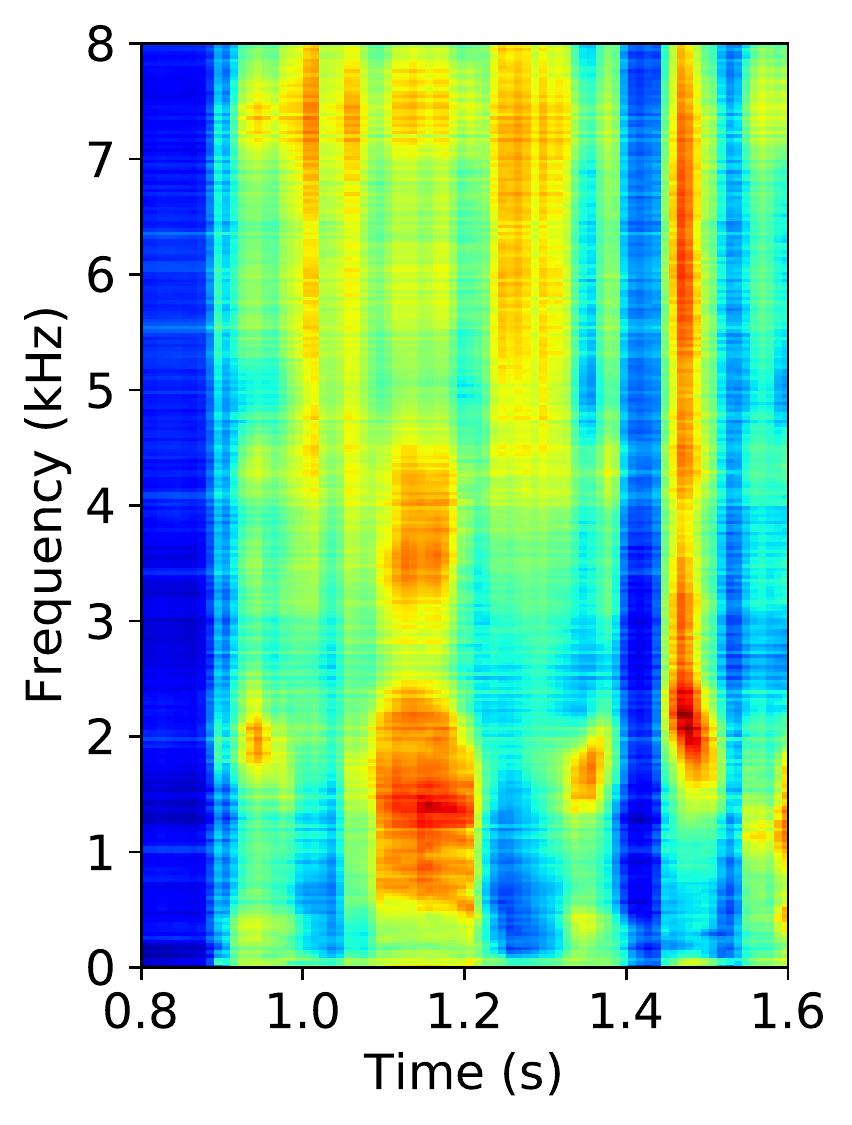}%
  \label{fig:evaluation:dnnjoint}%
}\\
\subfloat[Clean utterance]{%
  \includegraphics[width=6cm]{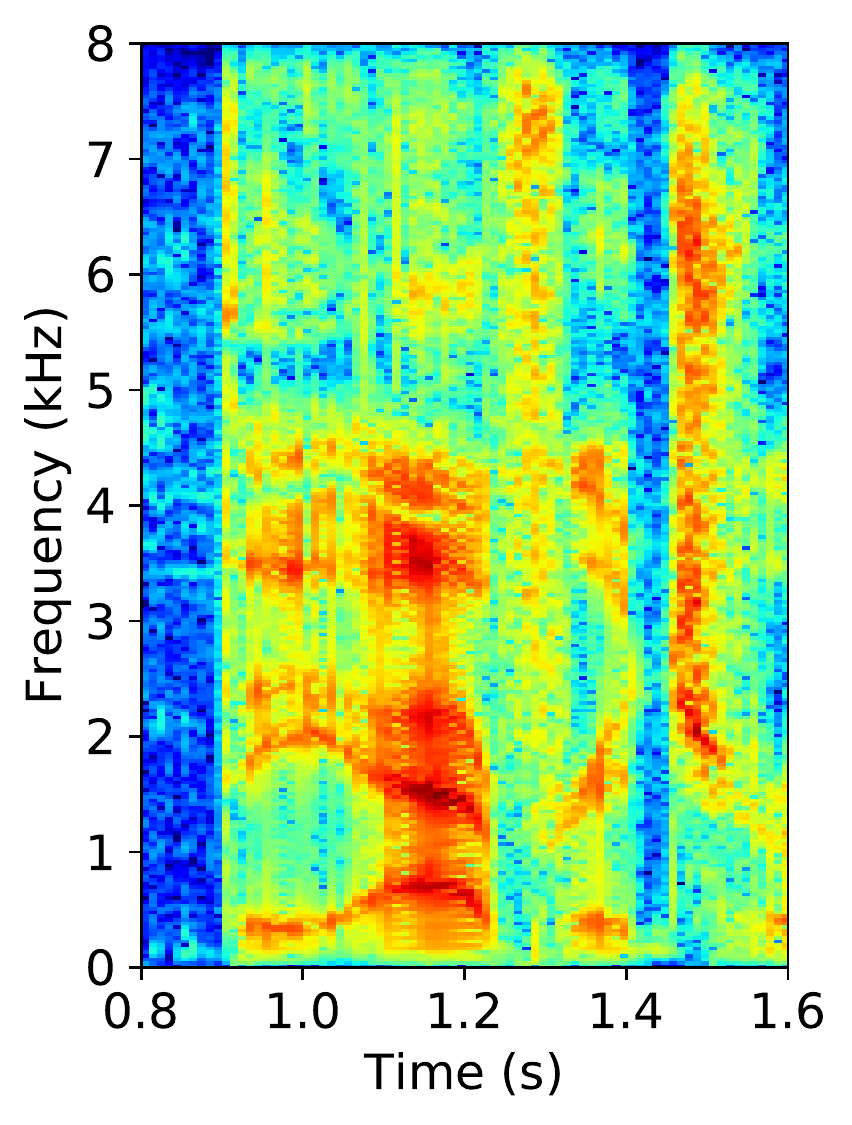}%
  \label{fig:evaluation:clean}%
}%
\subfloat[Wide ResNet mapper w/ L1 loss]{%
  \includegraphics[width=6cm]{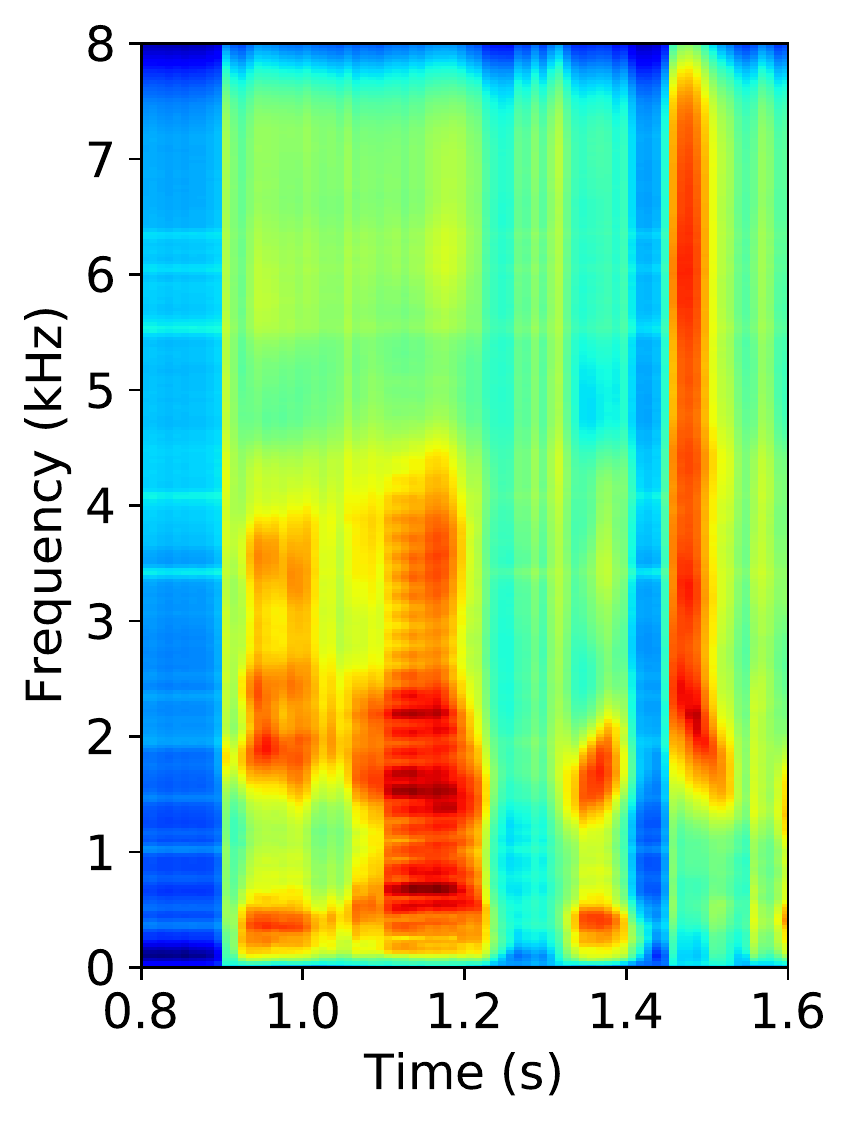}%
  \label{fig:evaluation:resnetfidelity}%
}%
\subfloat[Wide ResNet + perceptual Wide ResNet]{%
  \includegraphics[width=6cm]{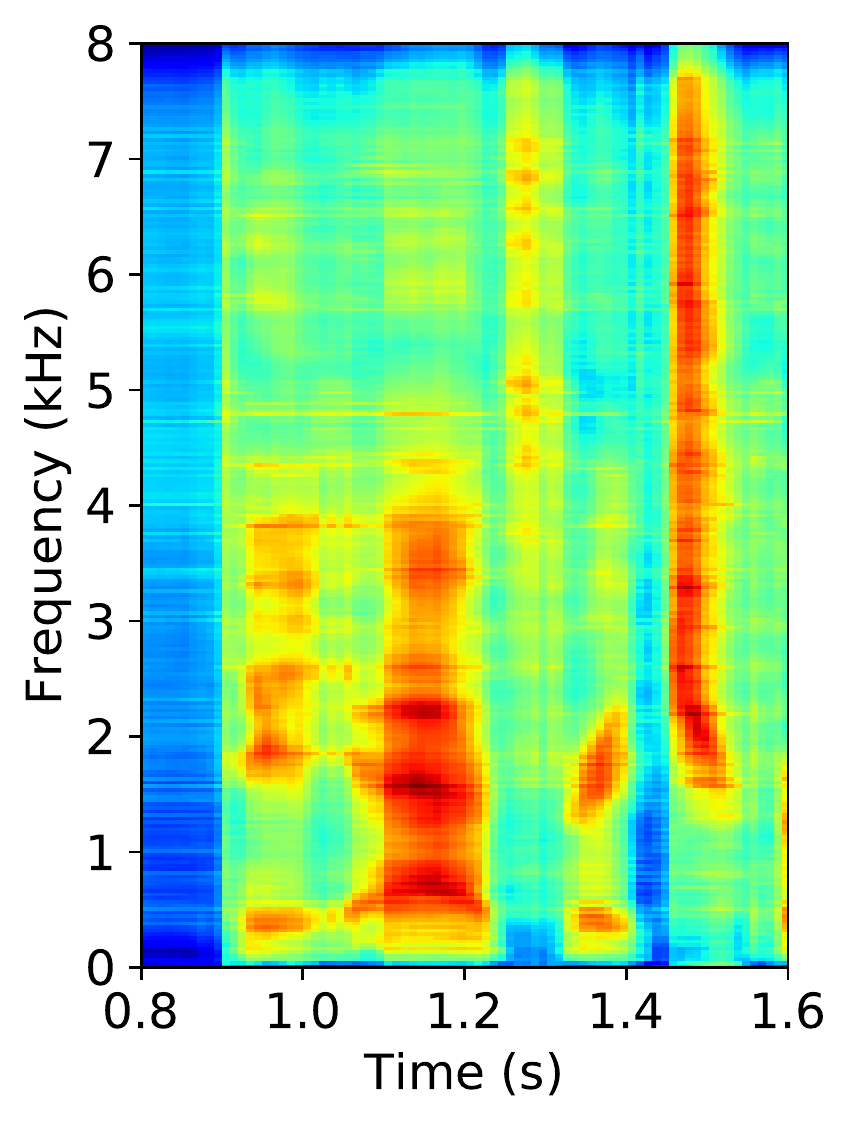}%
  \label{fig:evaluation:resnetjoint}%
}
\vspace{0.5cm}
\caption{We compare features for utterance 440c020f7 at -6dB SNR. Correct transcript is
    ``The average rate...'' See phoneme
predictions in Figs.~\ref{fig:posterior} and \ref{fig:likelihood}.
}
\label{fig:comparison}
\end{figure*}
\begin{figure*}
    \subfloat[Standard loss posteriorgram \\ (alignment: SIL D IH Z AE S T ER T R)]{
        \includegraphics[width=8.8cm,height=4.5cm]{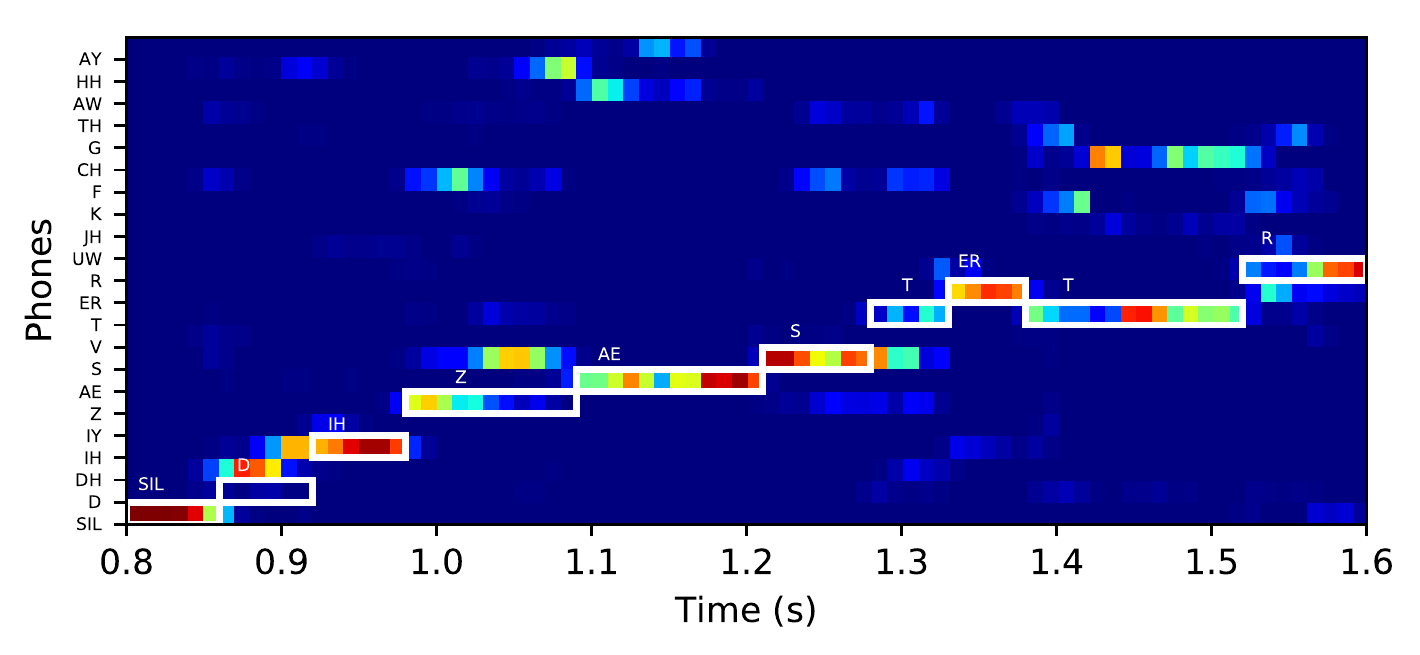}
        \label{fig:posterior:fidelity}
    }
    \subfloat[Perceptual loss posteriorgram \\ (alignment: SIL DH IY AE V R IH JH R)]{
        \includegraphics[width=8.8cm,height=4.5cm]{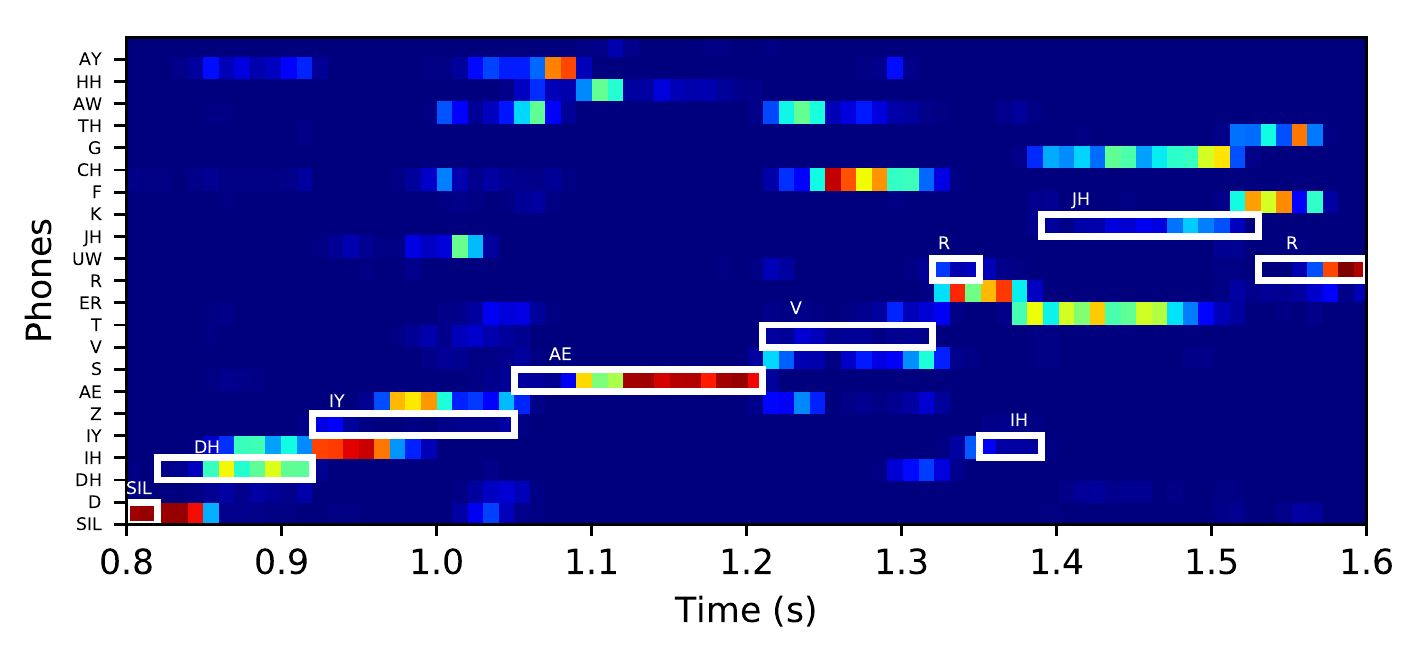}
        \label{fig:posterior:mimic}
    }
    \caption{Posterior comparison of L1 spectral loss and perceptual loss
    for utterance 440c020f7 ``The average rate...'' which is recognized correctly with perceptual loss, but the L1 spectral loss system recognizes as ''Disaster trade....''. Some phonemes left out due to space constraints.
    At around 1.3 seconds, the probability of /S/
    has been shifted to /V/ and /F/. See Fig.~\ref{fig:likelihood}
    for likelihood estimation.}
    \label{fig:posterior}
\end{figure*}
\begin{figure*}
    \subfloat[Standard loss likelihoodgram]{
        \includegraphics[width=8.8cm,height=4.5cm]{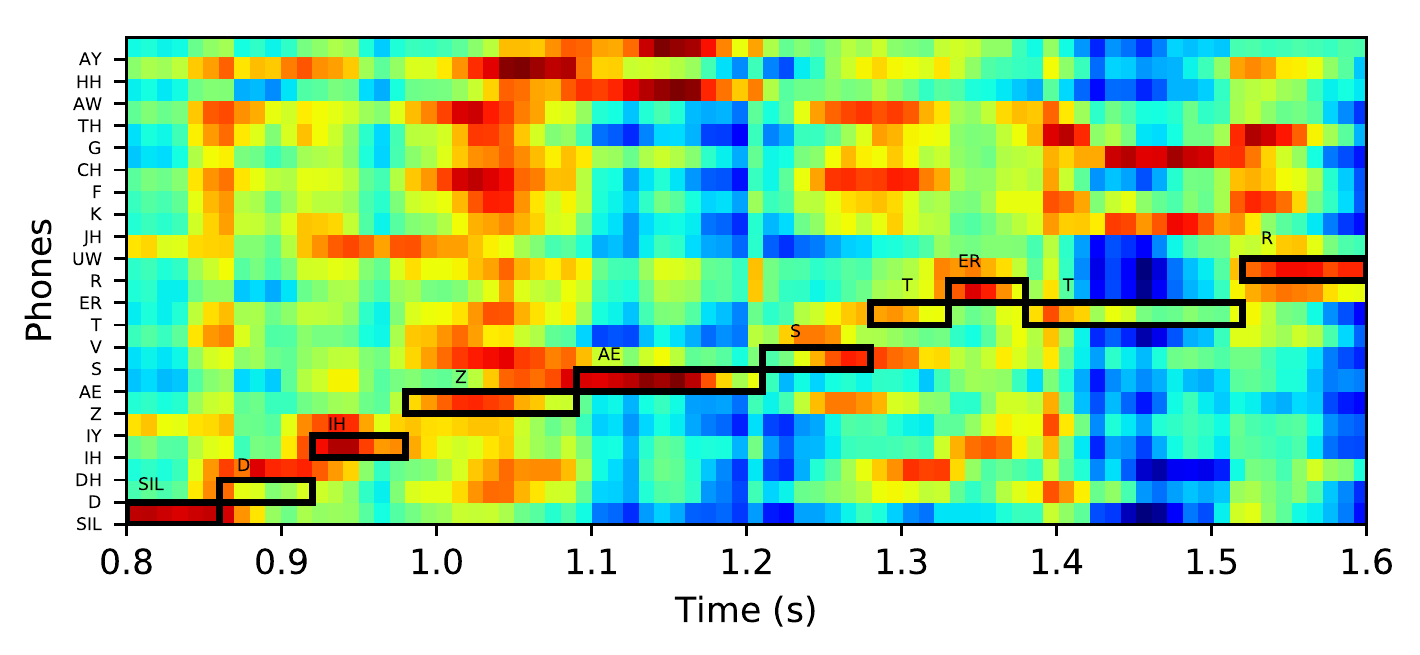}
        \label{fig:likelihood:fidelity}
    }
    \subfloat[Perceptual loss likelihoodgram]{
        \includegraphics[width=8.8cm,height=4.5cm]{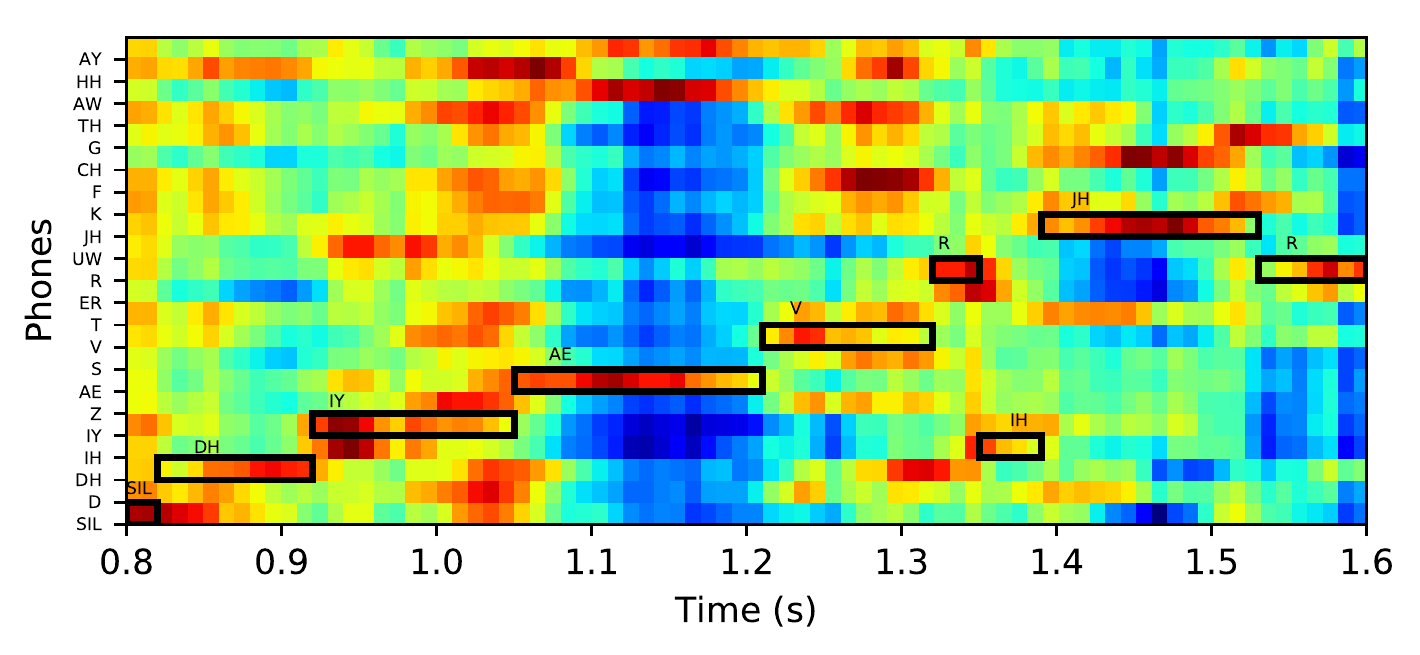}
        \label{fig:likelihood:mimic}
    }
    \caption{Likelihood comparison for utterance 440c020f7 ``The average rate...'' which is recognized correctly with perceptual loss, but the L1 spectral loss system recognizes as ''Disaster trade....''
    Some phonemes left out due to space constraints. See
    Fig.~\ref{fig:posterior} for posterior probabilities.}
    \label{fig:likelihood}
\end{figure*}

We first report the results of our follow-up experiments in Table~\ref{tab:chime2}.
The starred results in this table
are newly reported experiments, conducted in 2018 shortly after the deadline for
our submission~\cite{plantinga2018exploration}.

While our hypothesis was that
better recognizers should provide better feedback than worse recognizers, these
results demonstrate that this is not always the case. The WRBN model achieves almost
half of the cross-entropy loss relative to Wide ResNet, but is not as effective at
perceptual feedback. We think this is an important point overlooked so far in
literature regarding perceptual losses. We go into more detail and explore several
variations in Section~\ref{ssec:perceptual_training}. We suspect the fact that
the DNN model did not benefit from the Wide ResNet perceptual model is due to
having reached the maximum modeling capacity for the DNN.

In this table we also compare against the state-of-the-art for CHiME-2. The best-performing
recognition model on the CHiME-2 dataset is the WRBN recognition model trained with
distortion-independent training and a Gated Residual Network (GRN) enhancement front end~\cite{wang2019bridging}.
This work was done
in 2019 after our work was completed, though we haven't reported our results until now.
Our model tends to perform better in less noisy situations, perhaps due to introducing
fewer distortions that harm recognition performance.

To analyze the results further, we experimented with using different outputs from
the DNN perceptual model for generating the perceptual loss. The results in 
Table~\ref{tab:layers} show that in a context-constrained model, using deeper 
representations tend to generate better results. However, a large benefit is seen
even using just
a single layer. This could be helpful in resource-constrained environments.
We did not find that deeper outputs are always better, as revealed 
by our Voicebank + DEMAND
results discussed in the following sections.

In addition to our follow-up experiments, we performed an in-depth analysis of our
perceptual loss to see where it may help the most. First, 
we examined a specific utterance where the recognition score
improved due to the addition of perceptual loss. The utterance label is 440c020f7
and the transcript starts with ``The average rate is...''. Without
perceptual loss, this utterance's predicted transcription was ``Disaster trade...''.

In Fig.~\ref{fig:comparison}, we can see a comparison of features
generated by various systems for enhancement. The DNN model is less suitable for
enhancement, producing ``spectral leakage'' or artificial high-frequency noise.
This noise is greatly suppressed with perceptual loss. In the Wide ResNet case, no phoneme is generated at 1.3 seconds. With perceptual loss, there is a definite phoneme at 1.3 seconds.

The better reproduction results in a better prediction, as seen in Figs.~\ref{fig:posterior} and~\ref{fig:likelihood}. Instead of a strong prediction of /S/, the model shifts its confidence to /F/ and /V/. The shift to /F/ is visible in both posteriorgram and likelihoodgram but the emphasis on /V/ is only seen in the likelihoodgram, near 1.25 seconds.

Our hypothesis is that energy-based metrics can sometimes overlook low-energy phonemes such as /V/. By contrast, a phonetic perceptual loss penalizes the enhancement model for failing to reproduce phonemes regardless of their energy level. We expected perceptual loss to improve performance on low-energy phonemes more than high-energy phonemes.

In order to find evidence to support our hypothesis, we extend our analysis from one example to the entire dataset. First, we computed the phone sequences corresponding to the predicted words and compared them against the correct phone sequences. We computed the precision for each phoneme using the predictions of the Wide ResNet model when trained both with and without perceptual loss. We plotted the improvement in precision due to the addition of perceptual loss as one variable.

\begin{figure}
    \centering
    \includegraphics[width=\linewidth]{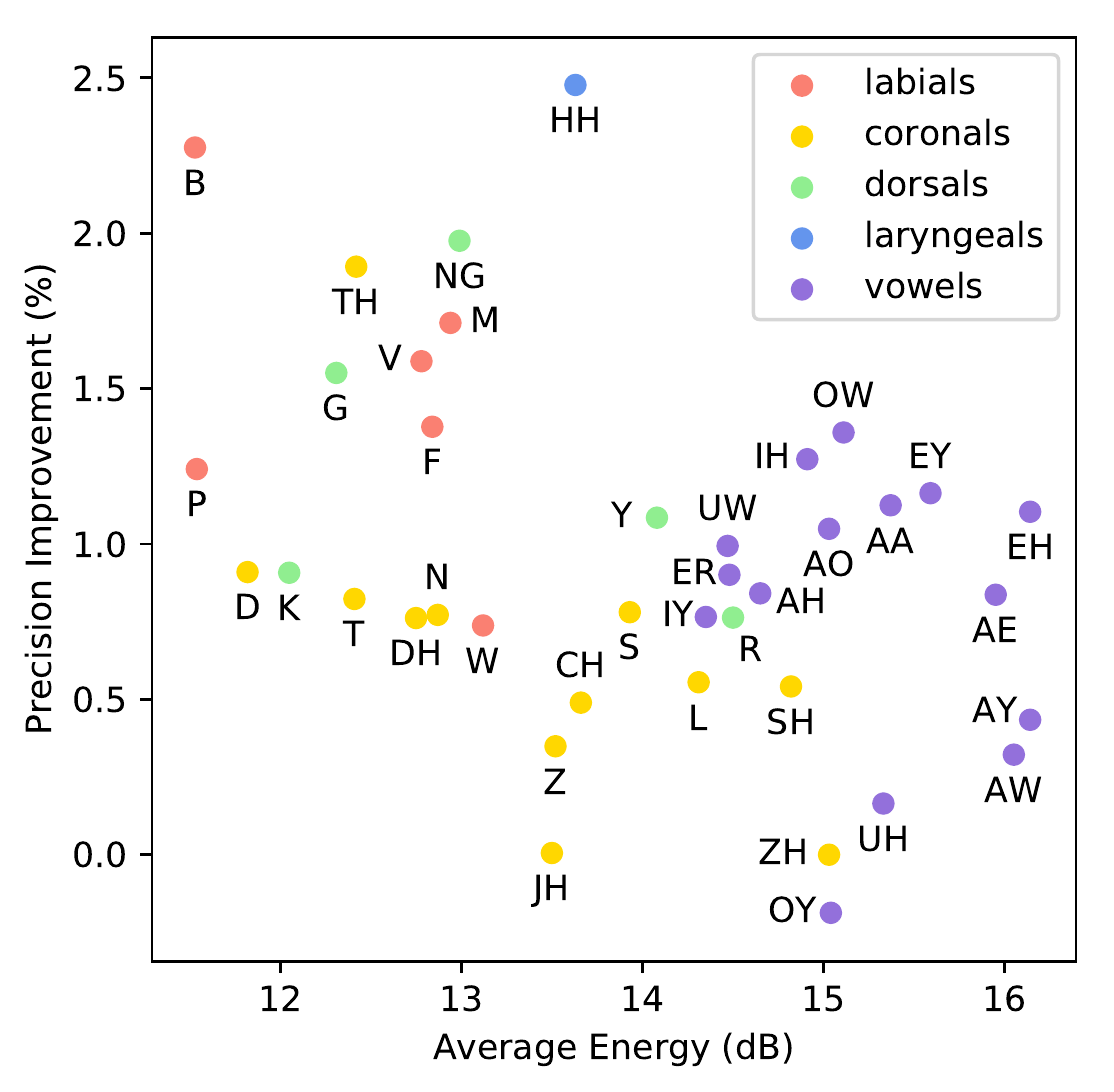}
    \caption{Average Phoneme Energy vs. Phoneme Precision Improvement}
    \label{fig:energy_vs_precision}
\end{figure}

Second, we computed the average energy of each phoneme across the training set. Specifically, we computed the average of the spectral frames selected using the senone alignments generated for acoustic model training. We plotted this variable against the precision improvement computed in the previous step.
The result is in Fig.~\ref{fig:energy_vs_precision}.

Taken as a whole, the figure shows a small but significant correlation between precision
improvement and average energy. The Pearson's correlation coefficient is -0.43 for all phonemes. When limiting the set of phonemes to consonants only the coefficient is -0.49, indicating that the trend is more apparent in lower-energy phonemes. This provides evidence supporting our hypothesis that perceptual loss improves the reproduction of low-energy phonemes more than high-energy phonemes.

\section{Perceptual vs. Joint Training \\ on the Voicebank + DEMAND corpus}
\label{sec:voicebank_procedure}

In addition to our CHiME-2 results and analysis, we report a new set of experiments
done with the SpeechBrain~\cite{ravanelli2021speechbrain}
toolkit. By contrast with Kaldi, this toolkit allows us to combine
state-of-the-art enhancement and end-to-end
recognition models in a straightforward way. We report for the
first time results directly comparing joint training against
perceptual loss training.

We evaluate our approach on the publicly-available Voicebank + DEMAND
corpus~\cite{valentini2016investigating}, and make our recipe publicly
available as a part of the SpeechBrain toolkit\footnote{\url{https://github.com/speechbrain/speechbrain/tree/develop/recipes/Voicebank/MTL/ASR_enhance}}. Our recognition results
are generated by fine-tuning a good-performing recognition system
trained on the LibriSpeech dataset. In the joint training scenario,
the weights of the enhancement model are updated during fine-tuning.

In addition to our comparison with joint training,
we experiment with ways of training perceptual
models in an end-to-end manner. The standard approach for ASR
in SpeechBrain is sequence-to-sequence training
with an encoder-decoder model and both connectionist temporal classification (CTC)~\cite{graves2006connectionist}
and negative log likelihood (NLL) losses.

After the perceptual model is trained, we use this model to generate
feedback during enhancement model training. Both clean and
denoised features are passed through the perceptual model,
and a loss is generated at the outputs. We combine L1 loss
at the outputs with standard L1 loss at the feature level.

Finally, after enhancement model is trained, our proposed approach
is to freeze the weights and fine-tune a large ASR model
pre-trained on LibriSpeech. We compare against a joint training
baseline where the weights of the enhancement model are allowed
to update during ASR fine-tuning.

More details on these three steps are provided in the following sub-sections.

\subsection{Perceptual Model Training}
\label{ssec:perceptual_training}

Central to our good results for enhancement and robust ASR is the careful training
of the perceptual model. As we demonstrated with our CHiME-2 experiments, better
recognizers do not always correspond to better perceptual models. We explore a
variety of factors to determine the most effective settings for a perceptual
model, as judged by objective enhancement scores. The scores we use are the
widely-used PESQ~\cite{pesq}, as well as the composite measure COVL~\cite{hu2007evaluation} that was shown to match human judgements very closely.

Before we relate the experimental settings, we first report the common settings.
The spectral features we use are 257-dimensional, generated with 32ms windows
and 16ms hop. These features are slightly worse for recognition than 25ms+10ms
features that are common for ASR, but are much better for enhancement. We also
use the log of the magnitude plus 1 to reduce the effect of very small 
or very large differences.

All of the perceptual models are trained in a sequence-to-sequence setup, using
a 1-layer 256-unit gated recurrent unit (GRU) model as the decoder. The loss is standard NLL loss against the targets. In the first few epochs of training we add an
output layer and CTC loss to the encoder model, as described in~\cite{kim2017joint}.

However, we find that our sequence-to-sequence models do not always converge,
likely due to the small size of the Voicebank dataset (8.8 hours of audio)
relative to other ASR datasets (100s or 1000s of hours). This
is especially a problem with our non-recurrent encoder model
(i.e. Wide ResNet). We address this with the
addition of \textit{alignment loss}~\cite{plantinga2019towards}.

When CTC loss fails to converge, the model tends to always output
the blank label. One way to counteract this tendency is to
encourage the model to produce non-blank outputs when someone
is speaking. The alignment loss term does this with a classification-like
loss (negative log likelihood) with the categories of silence or non-silence.
For our experiments we fixed the original incorrect formulation
which used positive log and swapped the silence and non-silence indicator terms.

Our indicator function separates frames by whether
they constitute silence or non-silence. The function is 1 for
blank outputs during silence and non-blank outputs during non-silence.
To be precise, if $\bar{E}$ is the average energy,
and $t$ represents the current frame, and $n$ is the index
of a phoneme, our indicator function is:

\begin{equation}
    I(t, n) = \begin{dcases}
        1 & \text{if } E_t < \bar{E} \text{ and } n = \epsilon \\
        1 & \text{if } E_t > \bar{E} \text{ and } n \neq \epsilon \\
        0 & \text{if } E_t < \bar{E} \text{ and } n \neq \epsilon \\
        0 & \text{if } E_t > \bar{E} \text{ and } n = \epsilon \\
    \end{dcases}
\end{equation}

Then our loss term is simply the negative log of the relevant portion of the posterior, selected with the indicator function:

\begin{equation}
    L_{align}(t) = -\log(\sum_{n=1}^N I(t, n) P_n)
\end{equation}

\begin{table}
    \centering
    \caption{Objective enhancement scores on Voicebank + DEMAND for two
    perceptual models, output at various layers.}
    \label{tab:context}
    \begin{tabular}{|l|c|cc|}
        \toprule
        Context model & Context width & PESQ & COVL \\
         \midrule
        Wide ResNet, first block & 7 frames & 3.03 & 3.74 \\
        Wide ResNet, second block & 13 frames & 2.97 & 3.68 \\
        CRDNN, first CNN block & 5 frames & 2.93 & 3.64 \\
        CRDNN, 2 CNN blocks & 9 frames & 3.05 & 3.74 \\
        CRDNN, 3 CNN blocks & 13 frames & \textbf{3.06} & \textbf{3.75} \\
        CRDNN, RNN outputs & full utterance & 2.91 & 3.54 \\
        CRDNN, DNN outputs & full utterance & 2.87 & 3.50 \\
         \bottomrule
    \end{tabular}
\end{table}

\begin{table}
    \centering
    \caption{Objective enhancement scores on Voicebank + DEMAND
    using different targets for training the perceptual model.
    Error rates are not comparable, but still
    indicate a variety of recognition abilities.}
    \label{tab:targets}
    \begin{tabular}{|l|c|cc|}
        \toprule
        Targets & Error Rate & PESQ & COVL\\
        \midrule
        characters & CER 20.37 & 2.98 & 3.68 \\
        word pieces & WER 21.19 & 2.92 & 3.62 \\
        phonemes & PER 14.55 & \textbf{3.05} & \textbf{3.74} \\
        \bottomrule
    \end{tabular}
\end{table}

\begin{table}
    \centering
    \caption{Objective enhancement scores on Voicebank + DEMAND using
    perceptual models trained with different inputs. Recognition rates
    included to indicate variety of recognition abilities.}
    \label{tab:inputs}
    \begin{tabular}{|l|c|cc|}
        \toprule
        Inputs & PER & PESQ & COVL \\
        \midrule
        Spectral magnitude & 14.55 & \textbf{3.05} & \textbf{3.74} \\
        Log-mel Filterbank (n=40) & 20.25 & 3.03 & 3.72 \\
        Log-mel Filterbank (n=80) & 15.36 & 3.02 & 3.73 \\
        \bottomrule
    \end{tabular}
\end{table}

Next, we turn to the experimental factors that we explore for our perceptual model. Primarily we explore three factors:

\begin{enumerate}
    \item Effect of context (changing model type and output layer)
    \item Effect of various training targets (e.g. characters or phonemes)
    \item Effect of various input types (e.g. spectrogram or filterbank)
\end{enumerate}

When training an acoustic model to achieve the best recognition performance, some
or all of the choices you make for these three factors may be different. We find that
recognition scores should not be used guide selection of perceptual models, other
than to ensure the model has converged.

The first factor that we experiment with is the context that is provided to the
acoustic model. The results of this experiment are shown in Table~\ref{tab:context}.
We experiment both with a Wide ResNet-like model, as in our CHiME-2 experiments, as
well as a CRDNN model that has shown good performance for recognition. These encoder
models are trained with the same decoder model, a 1-layer GRU with 256 units.
In both cases we find that deeper layers are not as effective as shallower layers.

In the case of Wide ResNet, our procedure is slightly different from the CHiME-2
work, as we trained the model in an utterance-wise fashion rather than
passing frame + context. This approach is much faster at both training and inference
time. Using this approach means that deeper layers have access to progressively
more context, and we find that this is not helpful for perceptual loss.
In the case of CRDNN, deeper CNN layers do seem to help, but the RNN and DNN
outputs are not effective for perceptual loss.

While at first these results seems to contradict what we found with the DNN model for
CHiME-2, we conjecture that the addition of more context hurts results.
Perhaps embeddings that only see a small context window are more focused
in their feedback.
Whatever the case, these results confirm the results we
got on CHiME-2 that suggest
that generating perceptual loss after recurrent layers is less effective than
only after non-recurrent layers.

The second factor that we consider is the targets. We find that training the perceptual model with character or word targets is much less effective than using phonetic targets. Results can be seen in Table~\ref{tab:targets}.
This is a clear indication that important phonetic information is being
captured by the perceptual model and conveyed to the enhancement module. One more note is that even with alignment loss the ASR model did not converge as well with word piece targets. Our first result was a WER of around 80\%. We reduced the output size using max pooling over 4 frames to reach a WER of 21\%.

The last factor we explored was the inputs. We report results with commonly used log-mel filterbank
features with both 40 and 80 filters in Table~\ref{tab:inputs}.
The pretrained model's performance varies a little with input size (20.3 vs 15.4)
but this does not seem to have much of an impact on the quality of the model
when used as a perceptual model.

This exploration shows that it is not always the best-performing features using
standard ASR metrics that provide the best feedback to an enhancement model.

\subsection{Enhancement Training}
\label{ssec:enhancement}

The model that we use for enhancement is based on the Wide-ResNet-like model presented
in~\cite{plantinga2018exploration}. In that work we motivate using a residual structure
for the enhancement model by noting its similarity to the task at hand. Since noisy
speech is usually modeled as (and artificially created as) a sum of clean and noise
signals, the enhancement task is to reproduce the input speech signal (i.e. residual)
without the added noise.

We made several updates to the residual network structure based mainly on new work that
has come out since we first used Wide ResNets.
The updates we made are as follows:

\begin{enumerate}
    \item The largest gain came from adding squeeze-and-ex\-citation
    blocks~\cite{hu2018squeeze} to the residual computation.
    \item Instead of performing spectral mapping to predict
    features directly, we used the spectral approximation
    algorithm~\cite{liu2019supervised}, applying outputs
    to noisy inputs as a mask and generating an L1 loss against clean targets.
    \item We added 2d batch normalization after each activation.
    \item We made the model deeper, adding two 512 filter blocks.
    \item We gained a small amount from changing the activation function to
    GELUs~\cite{hendrycks2016gaussian}.
    \item Our results showed another small gain (around 0.02 PESQ)
    from applying our outputs as a mask to the complex spectrogram,
    while still computing the loss using the log spectral magnitude (with no added RI loss).
    Enhancement model outputs are real-valued, clamped between 0 and 1, and then applied
    to the complex valued noisy spectrogram as a mask.
\end{enumerate}

Beyond our updates to the model, we find another small improvement from using L1 loss
rather than L2 loss for both spectral and perceptual losses.
We maintain the structure of adding a scaling constant $\alpha$
used for balancing the magnitude of the two losses.

\subsection{ASR Fine-tuning}
\label{ssec:fine_tuning}

In order to generate state-of-the-art recognition rates on the Voicebank + DEMAND
corpus, we fine-tune a large ASR system trained on LibriSpeech. This model targets
word pieces and includes a language model and beam search decoding for a final performance of
3.0 on LibriSpeech's test-clean.

During fine-tuning, we address the ``distortion problem'' by adding noises from the OpenRIR data set~\cite{ko2017study} at a random level between 0 and 15 dB. This is a standard data augmentation method in SpeechBrain, a part of the environmental corruption set of augmentations. This method helps both joint training and independent training, but helps independent training more than joint training. This is because the frozen enhancement models produce distortions when they encounter unseen noises, forcing the recognition model to adapt during training.

We fine-tune this model on Voicebank + DEMAND for 30 epochs, decreasing the learning
rate every 3 epochs by a factor of 0.7. Crucial to our results, we freeze the enhancement
and encoder model for the first three epochs, giving the decoder a chance to adapt
to language differences before the encoder model adaptation begins. Without this ``slow thaw'' approach, we saw more variability in our results and a higher average WER.

\section{Voicebank + DEMAND Results}
\label{sec:voicebank_results}

Our first set of results in Table~\ref{tab:mimic_results} show side-by-side
both objective enhancement scores and recognition scores, comparing
joint training against perceptual training. The jointly trained systems
perform worse than the systems trained with distortion-independent training,
confirming the research in~\cite{wang2019bridging}. In addition, the
perceptual loss improves both enhancement scores and test-set recognition
performance.

\begin{table}
    \centering
    \setlength{\tabcolsep}{5pt}
    \caption{Direct comparison of joint training and distortion-independent
    training on Voicebank + DEMAND. Both objective enhancement scores and
    recognition rates are recorded for the same model.}
    \label{tab:mimic_results}
    \begin{tabular}{|lcc|cc|cc|}
        \toprule
         & Enhancement & Joint & & & dev & test \\
        System & training loss & training & PESQ & COVL & WER & WER \\
        \midrule
        Clean & - & - & 4.50 & 4.50 & 1.44 & 2.29 \\
        Noisy & - & - & 1.97 & 2.63 & 4.32 & 3.43 \\
        Mask & L1 Spectral & Yes & 2.46 & 3.32 & 3.12 & 3.77 \\
        Mask & + L1 Perceptual & Yes & 2.44 & 3.29 & 3.57 & 3.58 \\
        Mask & L1 Spectral & No & 2.99 & 3.69 & \textbf{2.88} & 3.25 \\
        Mask & + L1 Perceptual & No & \textbf{3.05} & \textbf{3.74} & 2.89 & \textbf{2.80} \\
        \bottomrule
    \end{tabular}
\end{table}

We generate an SNR breakdown of the results by combining the development and test sets. The results are shown in Table~\ref{tab:snr_breakdown}, where it can be seen that adding perceptual loss improves recognition rates over the direct masking approach in all but one category. One interesting note is that perceptual loss maintains 25-30\% relative improvement over the noisy model at all SNR levels, whereas the improvement from the L1 loss trained model decreases as SNR increases. At the lowest SNR levels, the L1 trained model has only an 8\% relative improvement. One possible interpretation is that the perceptual system produces fewer distortions that harm recognition in the environment where there is little noise to begin with. At the highest SNR level, perceptual loss actually matches the performance of clean speech, suggesting that it can produce outputs with few or no distortions that harm recognition.

\begin{table}
    \centering
    \setlength{\tabcolsep}{4pt}
    \caption{WER scores for Voicebank + DEMAND dataset, broken down by SNR. Dev
    and test sets are grouped together for more stable results.}
    \label{tab:snr_breakdown}
    \begin{tabular}{|lc|cccc|}
        \toprule
        System & Enhancement loss & 0-3 dB & 5-8 dB & 10-13 dB & 15-18 dB \\
        \midrule
        Noisy & - & 7.53 & 3.31 & 2.07 & 2.64 \\
        Mask & L1 Spectral & 5.48 & \textbf{2.32} & 1.77 & 2.44 \\
        Mask & + L1 Perceptual & \textbf{5.26} & 2.45 & \textbf{1.55} & \textbf{1.99} \\
        Clean & - & 2.04 & 1.85 & 1.26 & 2.08 \\
        \bottomrule
    \end{tabular}
\end{table}

In addition to our comparisons with joint training and distortion-independent
training, we compare against other state-of-the-art systems on the Voicebank + DEMAND
dataset in Table~\ref{tab:voicebank_compare}. We note that our system performs best on CSIG and COVL, both highly
correlated with human judgements of signal quality. PHASEN performs better on
the CBAK measure, perhaps due to the fact that our system focuses on phoneme
reconstruction at the expense of ignoring background suppression. On the PESQ metric,
MetricGAN+ performs better than our system due to directly optimizing for a good PESQ score.
We have not included results from~\cite{kataria2021perceptual} which performs better than our system,
but is trained on additional data and is therefore not directly comparable.

\begin{table}
    \centering
    \caption{Comparison of objective enhancement scores for state-of-the-art systems
    on Voicebank + DEMAND.}
    \label{tab:voicebank_compare}
    \begin{tabular}{|ll|cccc|}
        \toprule
        System & Domain & PESQ & CSIG & CBAK & COVL \\
        \midrule
        Noisy & - & 1.97 & 3.35 & 2.44 & 2.63 \\
        SERGAN \cite{Baby2019SerganSE} & Time & 2.62 & - & - & - \\
        D. F. Loss \cite{MartnDoas2018ADL} & Time & - & 3.86 & 3.33 & 3.22 \\
        PHASEN \cite{yin2020phasen} & Complex & 2.99 & 4.21 & \textbf{3.55} & 3.62 \\
        DEMUCS \cite{defossez2020real} & Time & 3.07 & 4.31 & 3.40 & 3.63 \\
        MetricGAN+ \cite{fu2021metricganplus} & Spec. mag. & \textbf{3.15} & 4.14 & 3.16 & 3.64 \\
        \midrule
        Ours & Complex & 3.06 & \textbf{4.40} & 3.52 & \textbf{3.75} \\
        \bottomrule
    \end{tabular}
\end{table}

Finally, we test the performance of our enhancement system using a recognizer
trained on noisy data. The results are presented in Table~\ref{tab:no_adaptation}.
Since the recognizer is not adapted to enhanced outputs,
it is more seriously affected by introduced distortions. Still, with the addition
of perceptual loss the recognition rate improves, indicating fewer distortions.
More experimentation is needed to determine a cause for the worse performance
on the development set relative to training without perceptual loss.

\begin{table}
    \centering
    \caption{Recognition model trained on noisy utterances, evaluated
    using enhanced utterances.}
    \label{tab:no_adaptation}
    \begin{tabular}{|cc|cc|}
        \toprule
        ASR training data & Enhancement loss & dev WER & test WER \\
        \midrule
        Noisy & - & 4.32 & 3.43 \\
        Noisy & L1 Spectral & \textbf{3.36} & 3.58 \\
        Noisy & + L1 Perceptual & 3.69 & \textbf{3.34} \\
        \bottomrule
    \end{tabular}
\end{table}

\section{Conclusion}
\label{sec:conclusion}

We presented results showing that perceptual loss generated with
a recognition model can achieve state-of-the-art performance for
both noise-robust ASR and objective enhancement scores at the same time.

First, we reported follow-up experiments on CHiME-2 that matched the current state-of-the-art
recognition scores, performing better on higher-SNR samples. These experiments
demonstrated that better recognizers do not always make better perceptual models.
We also performed
an in-depth analysis of the CHiME-2 system to gain a better understanding of perceptual loss.
Our analysis shows that perceptual loss helps the most with low-energy
phonemes, due to the focus on phonemic structures rather than
myopically focusing on energy differences.

Second, we ran a new set of experiments on Voicebank + DEMAND in order
to directly compare against joint training. We confirm that distortion-independent
training indeed performs better than joint training, and find that
perceptual loss can recover some of the benefits of joint training while
avoiding the pitfalls. In addition, we experiment with different types
of perceptual models and find that phoneme targets, spectral magnitude
inputs, and small context windows are important qualities for
achieving good enhancement performance.

We released a recipe for perceptual loss training on Voicebank + DEMAND
using the SpeechBrain toolkit, as well as an interface for using a
pre-trained model using only a few lines of
code\footnote{\url{https://huggingface.co/speechbrain/mtl-mimic-voicebank}}.
This toolkit made our research possible
by providing good-performing recipes for ASR and enhancement that
were easy to combine for our purposes. We found the toolkit
to be an ideal resource for multi-task and multi-model work
due to the flexibility and modularity of the components.


%



\section*{Acknowledgment}
This work supported in part by the National Science Foundation
under Grant IIS-1409431 and Grant IIS-2008043.  We also thank the Ohio
Supercomputer Center (OSC)
\cite{OhioSupercomputerCenter1987} for providing us with computational
resources. We gratefully acknowledge the support of NVIDIA Corporation
with the donation of a Quadro P6000 GPU to our lab and a DGX-2 machine
to the SpeechBrain project, both used in this research.

Finally, we would like to thank Mirco Ravanelli, Szu-Wei Fu, Chien-Feng Liao, 
Ju-Chieh Chou, Abdel Heba, Aku Rouhe, Titouan Parcollet, Nauman Dawalatabad,
Samuele Cornell, and everyone else on the SpeechBrain team for their efforts
in creating the toolkit and developing recipes for VoiceBank + DEMAND
and LibriSpeech, as well as the pre-trained LibriSpeech seq2seq model.
\ifCLASSOPTIONcaptionsoff
  \newpage
\fi



\bibliographystyle{IEEEtran}
\bibliography{refs}
\end{document}